\documentclass[prl,twocolumn,showpacs,preprintnumbers,amsmath,amssymb,superscriptaddress]{revtex4}

\usepackage{graphicx}
\usepackage{dcolumn}
\usepackage{bm}
\usepackage{amsmath}

\begin{document}

\title{On Secure Quantum Key Distribution Using Continuous Variables of Single Photons}

\author{Lijian Zhang}
\email{l.zhang1@physics.ox.ac.uk}
  \affiliation{Clarendon Laboratory, University of Oxford, Parks Road, Oxford, OX1 3PU,United Kingdom}

\author{Christine Silberhorn}
     \affiliation{Institut f\"ur Optik, Information und Photonik,
     Universit\"at Erlangen-N\"urnberg, 91058 Erlangen, Germany}

\author{Ian A. Walmsley}
  \affiliation{Clarendon Laboratory, University of Oxford, Parks Road, Oxford, OX1 3PU,United Kingdom}

\begin{abstract}

We analyse the distribution of secure keys using quantum cryptography based on the continuous variable degree of freedom of entangled photon pairs. We derive the information capacity of a scheme based on the spatial entanglement of photons from a realistic source, and show that the standard measures of security known for quadrature-based continuous variable quantum cryptography (CV-QKD) are  inadequate. A specific simple eavesdropping attack is analysed to illuminate how secret information may be distilled well beyond the bounds of the usual CV-QKD measures. 

\end{abstract}

\maketitle



The distribution of secret information via optical channels, \textit{e.g.} quantum key distribution (QKD), provides an important example of the technological capability of  quantum correlations. The QKD protocol proposed by Bennett and Brassard~\cite{BB84} and its large collection of variations~\cite{QKD}, including QKDs using nonorthogonal states~\cite{B92} and entangled photons~\cite{Ekert91}, employ single photons or photon pairs to ensure secure information transfer between the source (Alice) and receiver (Bob). The quantum information associated with the single-photon states in these schemes is encoded as dichotomic variables, \textit{e.g.} in the polarization or relative phases of single-photon superposition states~\cite{Tittel2000}. Thus the maximum achievable information transfer rate is intrinsically limited to one bit per photon. A newer development of QKD utilizes continuous variable (CV) multi-photon systems~\cite{Ralph1999, Reid2000, Silb2002} where the amplitude and phase quadratures of coherent states~\cite{Grosshans2001,Grosshans2003} or squeezed states~\cite{Hillery2000, Cerf2001} serve as the information carriers. CV-QKD systems potentially enable higher key distribution rate. Recently, single-photon CV-QKD employing the position and momentum observables has been suggested as a means to increase the information transfer rate by coding more than one bit per photon. Compared to quadrature-based CV-QKD, single-photon CV-QKD eliminates the local oscillators required for homodyne detection and, as we will show, decouples the channel loss from the quantum correlations. Experimental implementations have demonstrated the feasibility of these schemes by utilizing  the spatial freedom of single photons~\cite{Walborn2006} or entangled photon pairs generated by parametric down-conversion (PDC)~\cite{Neves2005, Almeida05}. Yet, the security of such schemes has not been analysed, and as we show here, this is not a trivial extension of either BB84 or the conventional CV-QKD security proofs. 

In this Letter, we evaluate the potential of the spatial properties of PDC for QKD by considering a realistic PDC source as well as practical detectors and a lossy quantum channel. The analysis here also works for CV-QKD employing the correlations of time-frequency entangled photon pairs~\cite{Khan07}. Spatial correlations are, however, easier to manipulate with current technology, allowing more complete assessments of the channel security. In our analysis we derive the mutual information of the communicating parties from measurable postion-momentum correlations of PDC states and bound the information of a potential eavesdropper (Eve) by analyzing detected photocount statistics. Our results lie in the region between conventional dichotomic and continuous variable QKD and highlight the differences between these alternative approaches in terms of the experimental imperfections corrupting the secrecy of the key exchange. Our security analysis, which is mainly based on an intercept-resend eavesdropping strategy, indicates that single-photon CV-QKD gives increased secure bit rates per photon for intermediate channel losses. 


During the process of PDC, the pump photon with wave vector $\bm{k}_{p}$ spontaneously splits into two lower frequency (signal and idler) photons with wave vectors $\bm{k}_{s}$ and $\bm{k}_{i}$. The spatial and spectral properties of the photon pair are correlated by the material dispersion. In what follows it is assumed that the state is spectrally filtered such that the frequencies of signal and idler photons are restricted to $\omega_{s0} = \omega_{i0} = \omega_{p}/2$. The resulting two-photon state is 
\begin{equation}
\label{PDC_state}
\left|\Psi \right\rangle \approx
\left|\textrm{vac}\right\rangle + \mu \iint
d\bm{k_{s}^{\perp}}d\bm{k_{i}^{\perp}}
\,f(\bm{k_{s}^{\perp}};\bm{k_{i}^{\perp}})
\left|\bm{k_{s}^{\perp}};\bm{k_{i}^{\perp}}\right\rangle.
\end{equation}
For the practical PDC source, the down-converted modes are usually close to the longitudinal axis with the transverse vectors $|\bm{k}_{s}^{\perp}|\ll k_{s}$ and $|\bm{k}_{i}^{\perp}|\ll k_{i}$ ($k_{s/i}=|\bm{k}_{s/i}|$). For a pump beam with a Gaussian profile the  biphoton amplitude $f(\bm{k_{s}^{\perp}};\bm{k_{i}^{\perp}})$ can be approximated by
\begin{align}
f(\bm{k}_{s}^{\perp};\bm{k}_{i}^{\perp}) = \alpha (\bm{k}_{s}^{\perp}+\bm{k}_{i}^{\perp}) \phi_{L} (\bm{k}_{s}^{\perp}-\bm{k}_{i}^{\perp})\nonumber \\
= C \exp
\left[-\frac{w_{0}^{2}}{4}\left|\bm{k}_{s}^{\perp}+\bm{k}_{i}^{\perp}\right|^{2}\right]
\frac{\exp \left(i\Delta k_{z}L\right)-1}{i\Delta k_{z}L}, \label{joint_amplitude}\\
\textrm{with }\Delta k_{z}\approx 2K-k_{p}-\frac{\left|\bm{k}_{s}^{\perp}-\bm{k}_{i}^{\perp}\right|^2}{4K} \nonumber
\end{align}
and where $\alpha (\bm{k}_{s}^{\perp}+\bm{k}_{i}^{\perp})$ originates from the pump envelope and transverse phase-matching function, while $\phi_{L} (\bm{k}_{s}^{\perp}-\bm{k}_{i}^{\perp})$ is the longitudinal phase-matching function. $C$ is the constant for normalization,
$K=k_{s}=k_{i}$, $w_{0}$ is the beam waist of the pump and $L$ denotes  the length of the nonlinear crystal in $z$ direction~\cite{footnote1}. Retaining the longitutional phasematching function $\phi_{L} (\bm{k}_{s}^{\perp}-\bm{k}_{i}^{\perp})$ is critical to bounding the shared information from above~\cite{footnote2}.

\begin{figure}
 \includegraphics[height=.3\textwidth]{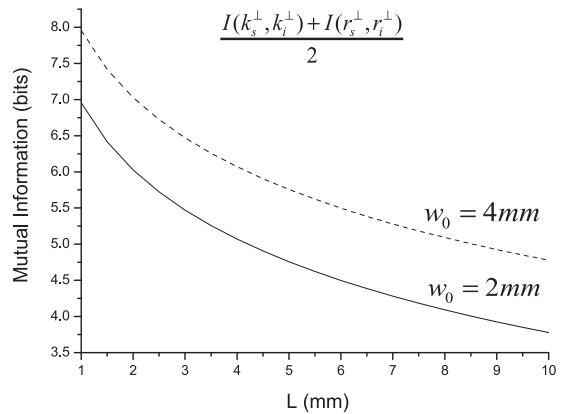}
 \caption{Mutual informations for entangled photon pairs generated by Type-I PDC in BBO crystal. This figure shows how the         crystal length and the pump waist affects the averaged mutual information in momentum and position.}
 \label{mutual_information_pic}
\end{figure}

The joint probability distribution of $\bm{k}_{s}^{\perp}$ and $\bm{k}_{i}^{\perp}$ is given by $p(\bm{k}_{s}^{\perp};\bm{k}_{i}^{\perp}) =
|f(\bm{k}_{s}^{\perp};\bm{k}_{i}^{\perp})|^{2}$. The mutual information between $\bm{k}_{s}^{\perp}$ and $\bm{k}_{i}^{\perp}$ can be calculated from~\cite{Shannon48}
\begin{equation}
\label{mutual_information}
I(\bm{k}_{s}^{\perp};\bm{k}_{i}^{\perp})=H(\bm{k}_{i}^{\perp})-H(\bm{k}_{i}^{\perp}\mid\bm{k}_{s}^{\perp}).
\end{equation}
where $H(\bm{k}_{i}^{\perp})$ and $H(\bm{k}_{i}^{\perp}\mid\bm{k}_{s}^{\perp})$ denote the entropy and conditional entropy respectively. Similarly, the Fourier transform of Eq. (\ref{joint_amplitude}) determines the mutual information $I(\bm{r}_{s}^{\perp};\bm{r}_{i}^{\perp})$ between the transverse positions of the two photons.
We model our practical source of entangled photon pairs by considering degenerate Type-I PDC in a BBO crystal with a phase-matching angle of $3^{\circ}$, pumped at 400nm.  Fig. \ref{mutual_information_pic} shows  the calculated maximum mutual information that Alice and Bob can extract if they adopt a symmetric coding, \textit{i.e.} they measure with equal probabilities the position and momentum of the photons. The graph illustrates the information transfer gain for CV single-photon systems. This should be compared with binary coding, for which a maximal value of one is obtained. For a fixed pump power, the amount of shared information between PDC photons
may be increased by increasing the pump waist $w_{0}$ and decreasing the crystal length $L$, though the penalty is a reduced efficiency of photon-pair generation, resulting in low signal rates. The entanglement of the two-photon state in our analysis can be characterized by considering the correlations or the mutual information for direct measurements of a pair of conjugate continuous variables, namely the position and momentum of the photons. Alternatively, one may quantify the entanglement contained in this degree of freedom by decomposing the state into its Schmidt modes~\cite{Law04},  and evaluating the corresponding concurrence.  We verified that
this approach yields the same asymptotic behavior, which confirms the consistency of our results with more general entanglement measures. QKD further requires that the measurements of non-corresponding variables do not exhibit correlations; our calculations show that the mutual information between momentum and position ($I(\bm{k}_{s}^{\perp};\bm{r}_{i}^{\perp})$ and $I(\bm{r}_{s}^{\perp};\bm{k}_{i}^{\perp})$) is negligible.

To analyse the security of a single-photon CV-QKD system we choose a specific protocol. Pairs of entangled photons are generated in the nonlinear crystal and transmitted to Alice and Bob separately via a quantum channel. The two parties choose randomly to detect either the position ($\bm{r}^{\perp}$) or momentum ($\bm{k}^{\perp}$) of each photon they receive. Then Alice and Bob announce by an authenticated public channel the variables which they measured for each photon and drop the bits where they used different variables; the remaining bits constitute the sifted raw key. To accomplish a successful quantum key distribution, the system must allow Alice and Bob to distill a secret key from the sifted raw key that is inaccessible to the adversary, Eve. With forward reconciliation~\cite{forward} and privacy amplification~\cite{privacy}, the achievable secret key rate in momentum is bounded below by
\begin{equation}
\label{secret_key}
\Delta I = I_{AB}-I_{AE} = H(\bm{k}_{A}\mid E) - H(\bm{k}_{A}\mid \bm{k}_{B}),
\end{equation}
where $E$ is the result of Eve's measurement on her ancilla. For individual attacks, it has been shown that there exists the entropic uncertainty relation~\cite{Grosshans2004}
\begin{equation}
\label{entropic_uncertainty}
H(\bm{k}_{A} \mid E)+H(\bm{r}_{A} \mid \bm{r}_{B}) \geq \log_{2}\pi e,
\end{equation}
The conditional entropy is bounded by~\cite{Cover_book}
\begin{equation}
\label{entropy_bound}
H(\bm{x}_{A}\mid \bm{x}_{B})\leq\frac{1}{2}\log_{2}\left[2\pi e\Delta^2(\bm{x}_{A}\mid \bm{x}_{B})\right],
\end{equation}
where $x$ stands for $k$ or $r$, while $\Delta^2$ denotes the variance. By combining Eqs. (\ref{secret_key}-\ref{entropy_bound}), we find
\begin{align}
\label{key_bound}
\Delta I &\geq \log_{2}\pi e - H(\bm{r}_{A} \mid \bm{r}_{B}) - H(\bm{k}_{A} \mid \bm{k}_{B})\nonumber \\    &\geq\frac{1}{2}\log_{2}\left(\frac{1}{4}\frac{1}{\Delta ^{2}(\bm{r}_{A}\mid \bm{r}_{B}) \Delta^{2}(\bm{k}_{A}\mid \bm{k}_{B})}\right).
\end{align}
So a sufficient condtion for $\Delta I \geq 0$ is 
\begin{equation}
\label{entanglement_witness}
\Delta ^{2}(r_{A} \mid r_{B}) \Delta^{2}(k_{A} \mid k_{B}) \leq \frac{1}{4}
\end{equation}
This result also applies for the security analysis in position. For high entanglement, this condition coincides with the EPR criterion~\cite{Reid1988}. It is easy to prove from Eq. (\ref{joint_amplitude}) that the states generated by PDC satisfy this condition, as demonstrated recently~\cite{Howell2004, Almeida05, Neves2005}. Note, however, almost all of these experiments employ one detector to scan through the momentum or position values, so in principle the outcome of each measurement is binary: either the photon hits the detector or not. Therefore this setup is not suitable for single-photon CV-QKD. To realise the full potential of continuous variables without complex encoding, a sufficiently large array of detectors (APDs, pixels of a CCD camera, etc.) is needed to ensure that binning and truncating do not significantly diminish the information transfer rate~\cite{footnote3}. This implies that the dark count of the detectors will have a much higher impact on the error rate than in standard BB84, though the probability that Eve can guess the correct result also decreases with the increased number of detectors.

To see this, assume that the entangled photon pair is generated from the pump pulse with probability $P_{PDC}$ and sent to Alice and Bob through two quantum channels with throughtputs $t_{A}$ and $t_{B}$. To measure the continuous variables $\bm{r}^{\perp}$ or $\bm{k}^{\perp}$ each party maps the distribution to $n$ identical detectors that are time-gated synchronously with the pump pulse. We denote the probability of recording a dark count within the detection time window for each detector as $P_{dark}$ and its efficiency as $\eta$. Alice and Bob keep the results when one and only one detector clicks. So there are three cases to be considered: (1) both parties have a dark count; (2) one party detects a photon and the other has a dark count; (3) both parties detect a photon. The probabilities for each case are:
\begin{subequations}
\label{prob_real}
\begin{align}
\begin{split}
P_{1}&=[1-P_{PDC}+ P_{PDC} (1-\eta t_{A})(1-\eta t_{B})] \\
&\quad\times n^2 P_{dark}^2(1-P_{dark})^{2n-2},
\end{split}\label{prob_1}\\
\begin{split}
P_{2} &= P_{PDC}[\eta t_{A}(1-\eta t_{B})+(1-\eta t_{A})\eta t_{B}] \\
&\quad\times n P_{dark} (1-P_{dark})^{2n-1},
\end{split}\label{prob_2}\\
P_{3} &= P_{PDC} \eta^{2} t_{A} t_{B} (1-P_{dark})^{2n} \label{prob_3}.
\end{align}
\end{subequations}
respectively. The probability that the photon and a dark count arise at the same detector simultaneously is negligible. Among all the cases, only $P_{3}$ will reveal the quantum correlations. This probability decreases as the channel loss and number of detectors increase due to the increase of the background noise level. Some typical values for the realistic system with APDs as detectors and nanosecond time gating are $P_{PDC}=0.01$, $\eta=0.6$ and $P_{dark}=10^{-6}$. We fix the length of the BBO crystal to 2mm and assume a 2mm pump waist (FWHM). The source lies at Alice's station, \textit{i.e.} $t_{A}\approx 1$ and $t_{B}=t$, where $t$ is the transmission of the channel between Alice and Bob. Taking into account the dark count contribution according to Eq. (\ref{prob_real}), Eq. (\ref{entanglement_witness})  is satisfied for channel throughput above $t=36\%$ ($68\%$) ($4.4$dB ($1.7$dB) channel loss) assuming a detector array with $n=128$ ($256$) pixels. For free space transmission the extinction coefficient varies over a large range~\cite{Freespace}. Here we assume it is $1$dB/km, so the corresponding distance is $4.4$km and $1.7$km respectively. At these distances the probability of uncorrelated events $P_{1}+P_{2}$ is less than 1\%, which means that the noise level is still extremely low.

Analysis of the variance product seems to suggest that this QKD scheme is not suitable for long-distance use. But we note that Eq. (\ref{entanglement_witness}) is a tight bound for general CV-QKD schemes and it is possible to loosen the bound when considering the special characteristics of the experimental imperfections in the single-photon CV-QKD protocol. Reconsidering  Eqs. (\ref{secret_key}-\ref{entanglement_witness}), note that the equality in Eq. (\ref{key_bound}) can only be achieved when Eve's attacks satisfy certain strict conditions. The most important condition is that the distribution of Bob's measurement outcomes conditioned on Alice's results should be Gaussian~\cite{Cover_book}. A Gaussian attack is well known to be optimal for conventional CV-QKD using the quadratures of multi-photon states since in these systems experimental imperfections---mainly the loss of the channel---will preserve the Gaussian character of the transmitted state, broadening Bob's distribution. By replacing the channel with a lossless one and applying a Gaussian attack, Eve can hide behind the existing experimental imperfections. The normal way for Alice and Bob to detect Eve is to monitor the covariance matrix of their results. In contrast, for  single-photon dichotomic-variable QKD, the experimental imperfections (loss, noise, etc) yield uncorrelated detection events between Alice and Bob, which are typically interpreted as background noise. In single-photon CV-QKD the experimental imperfections play a similar role to those in standard dichotomic single-photon QKD. The events registered by each party are either from the PDC photons or from the detector noise,  and the latter has a uniform distribution. Hence Alice and Bob expect un-broadened Gaussian joint probability distributions from the quantum correlation measurements  interspersed with uncorrelated flat background events, which in total represents a non-Gaussian distribution. In order to stay undetected Eve must mimic this distribution, therefore she only has limited options and the optimal attack for multi-photon CV-QKD is prohibited here. Moreover, for non-Gaussian distributions, the left side of Eq. (\ref{key_bound}) can be much bigger than the right side, which means even when the EPR condition is violated, it is still possible for Alice and Bob to draw the secret key.

\begin{figure}
 \includegraphics[height=.3\textwidth]{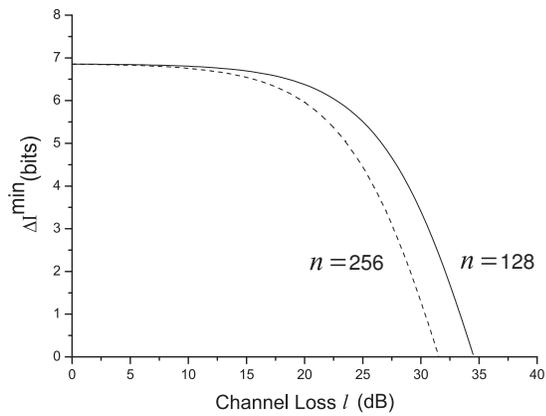}
 \caption{The minimum secret information per recorded photon pair ($\Delta I^{min}$) is estimated numerically from $\lambda_{max}$. The corresponding channel loss $l$ is calculated from Eq. (\ref{lambda_limit}). The initial entangled photons are generated by a 2mm-long BBO crystal and a pump at 400nm with 2mm beam waist.}
 \label{mutual_info_eve}
\end{figure}

A possible eavesdropping strategy that satisfies the above conditions is an intercept-resend attack: Eve intercepts the photon sent to Bob, measures it in the randomly chosen variable (momentum or position), and resends a photon in the eigenstate based on her measurement result. If, by chance, she has chosen the same measurement basis as Alice and Bob, her operation will appear as an undisturbed channel between these two parties. Otherwise, measuring the conjugate variable Eve introduces a flat background noise, which cannot be distinguished from the dark noise of the detector array. Therefore by adjusting the loss of the channel, Eve can hide her disturbance behind the experimental imperfections. We define an intercept-resend ratio $\lambda$  as
\[ \nonumber
\lambda = \frac{\textrm{Number of photons intercepted by Eve}}{\textrm{Total number of photons Alice sends to Bob}}.
\]
By balancing the disturbance introduced by Eve with the background noise, which originates from the experimental imperfections, we find an allowed maximum intercept-resend ratio for Eve is: 
\begin{equation}
\label{lambda_limit}
\lambda_{max}\approx\textrm{min}\left\{\frac{2n}{\left(\frac{1}{l}-1\right)\left(\frac{1}{P_{dark}}-1\right)+n}, \qquad 1\right\}
\end{equation}
where $l$ is the channel loss and $n$ is the number of detectors. For a lossless channel ($l=0$) or noiseless detectors ($P_{dark}=0$), $\lambda_{max}=0$, i.e. no eavesdropping is possible; while for fixed $l$ and $P_{dark}$, $\lambda_{max}$ increases with $n$. Eq. (\ref{lambda_limit}) clearly shows how the experimental imperfections open loopholes for Eve to attack. Moreover, the minimum secret information that Alice and Bob are able to distill ($\Delta I^{min}= I_{AB}^{min}-I_{AE}^{max}$) can be directly estimated from $\lambda_{max}$. The relation between $\Delta I^{min}$ and the channel transmission loss is shown in Fig. \ref{mutual_info_eve}. Comparing this result with the variance product analysis, it is evident that the secure loss level (35dB for $n=128$) is significantly improved for this eavesdropping strategy. 

An important question in quantum cryptography is the relationship between entanglement and security. It has been proved that distributed entanglement between Alice and Bob is a necessary precondition for secret key distribution~\cite{Curty04}. Also the connection between quantum and secret correlations has been established~\cite{Acin05}. Nevertheless it is still not clear how to draw a secure key from the distributed entanglement. For classical privacy amplification (forward or reverse reconciliation), the security limit is usually a stronger condition than the entanglement threshold~\cite{Grosshans2003_2}. In the intercept-resend attack for our protocol, the logarithm negativity as a function of the intercept fraction $\lambda$ shows that Alice and Bob remain entangled until $\lambda=1$, while as Fig. \ref{mutual_info_eve} and Eq. (\ref{lambda_limit}) show, the classical privacy amplification requires $\lambda < 75\%$ (where $\Delta I^{min}=0$) to draw the secret key. Hence for a practical QKD scheme, the detection of entanglement may not be enough for secret key distillation.  

To conclude, we have shown the potential to transfer more than one bit of information per photon using the spatial degrees of freedom of the entangled photon pairs. Due to the special non-Gaussian distributions of Alice and Bob's measurement results, the options for eavesdropping are severely limited. A detailed security analysis on a plausible attack, intercept-resend, is given. Whether Eve gains by means of more powerful attacks requires further study. In particular, a more detailed analysis of the impact of binning the information is required for a practical QKD system with limited number of detectors. Refinement of the security analysis will also take into account of the turbulence effects for free space transmission, which will give Eve more options to attack. 

This work was supported by EPSRC, EC under the Integrated Project QAP funded by the IST director as Contract No. 015848. L. Z. acknowledges support from the K.C. Wong Scholarship.



%




\begin{thebibliography}{Lijian06}
\bibitem{footnote2}  In previous treatements of this problem, $\phi_{L} (\bm{k}_{s}^{\perp}-\bm{k}_{i}^{\perp})$ has been ignored. See, for example, \cite{Almeida05}.
\bibitem{footnote3}  The number of detectors should satisfy $n\geq 2^{I_{AB}}$.
\end{thebibliography}
\end{document}